\RequirePackage{ifpdf}
\ifpdf 
\documentclass[pdftex]{sigma}
\else
\documentclass{sigma}
\fi

\def\tr{\operatorname{tr}}
\def\diag{\operatorname{diag}}

\def\gr{\operatorname{Gr}}
\def\Diag{\operatorname{Diag}}
\def\Span{\operatorname{span}}

\begin{document}
\allowdisplaybreaks

\renewcommand{\PaperNumber}{039}

\FirstPageHeading

\renewcommand{\thefootnote}{$\star$}

\ShortArticleName{Fine Gradings of Low-Rank Lie Algebras}

\ArticleName{Fine Gradings of Low-Rank Complex Lie Algebras\\ and of
Their Real Forms\footnote{This paper is a contribution to the Proceedings
of the Seventh International Conference ``Symmetry in Nonlinear
Mathematical Physics'' (June 24--30, 2007, Kyiv, Ukraine). The
full collection is available at
\href{http://www.emis.de/journals/SIGMA/symmetry2007.html}{http://www.emis.de/journals/SIGMA/symmetry2007.html}}}


\Author{Milena SVOBODOV\'A}

\AuthorNameForHeading{M.~Svobodov\'a}

\Address{Czech Technical University in Prague, Faculty of Nuclear
Sciences and Physical Engineering, Trojanova 13, 120 00, Praha 2,
Czech Republic}

\Email{\href{mailto:milenasvobodova@volny.cz}{milenasvobodova@volny.cz}}

\ArticleDates{Received August 31, 2007, in f\/inal form April
07, 2008; Published online April 14, 2008}


\Abstract{In this review paper, we treat the topic of f\/ine
gradings of Lie algebras. This concept is important not only for
investigating the structural properties of the algebras, but, on
top of that, the f\/ine gradings are often used as the starting
point for studying graded contractions or deformations of the
algebras.
One basic question tackled in the work is the relation between the
terms `grading' and `group grading'. Although these terms have
originally been claimed to coincide for simple Lie algebras, it
was revealed later that the proof of this assertion was incorrect.
Therefore, the crucial statements about one-to-one correspondence
between f\/ine gradings and MAD-groups had to be revised and
re-formulated for f\/ine group gradings instead. However, there is
still a hypothesis that the terms `grading' and `group grading'
coincide for simple complex Lie algebras.
We use the MAD-groups as the main tool for f\/inding f\/ine group
gradings of the complex Lie algebras $A_3 \cong D_3$, $B_2 \cong
C_2$, and $D_2$. Besides, we develop also other methods for
f\/inding the f\/ine (group) gradings. They are useful especially for
the real forms of the complex algebras, on which they deliver
richer results than the MAD-groups.
Systematic use is made of the faithful representations of the
three Lie algebras by $4\times 4$ matrices: $A_3 = sl(4,\mathbb
C)$, $C_2 = sp(4,\mathbb C)$, $D_2 = o(4,\mathbb C)$. The
inclusions $sl(4,\mathbb C)\supset sp(4,\mathbb C)$ and
$sl(4,\mathbb C) \supset o(4,\mathbb C)$ are important in our
presentation, since they allow to employ one of the methods which
considerably simplif\/ies the calculations when f\/inding the f\/ine
group gradings of the subalgebras $sp(4,\mathbb C)$ and
$o(4,\mathbb C)$.}


\Classification{17B45; 22E60}

\section{Introduction}\label{intro}

Gradings of Lie algebras have been explicitly used for more than
f\/ifty years. Probably the most notorious application is the
$\mathbb Z_2$-grading from the work of E.~In\"{o}n\"{u} and
E.~Wigner \cite{Inonu-Wigner}. Another well known fact is that
$\mathbb Z_2$-gradings play an important role for classif\/ication
of real forms of simple Lie algebras.

It was in 1989 that systematic studying of gradings of Lie
algebras has started in an article by J.~Patera and H.~Zassenhaus
\cite{OLG1}. In that article they have introduced the terms of
f\/ine grading and group grading, and investigated the role of
automorphisms for construction of gradings. A~number of works
followed \cite{Hav-4, formysl3c, OLG2, OLG3, Hav-3, HaPaPeTo},
using the theoretical results of that article for applications on
concrete Lie algebras. In recent years, gradings were intensively
studied not only on the classical f\/inite-dimensional Lie algebras
in \cite{Bahturin, bahturin-2}, but also on the exceptional Lie
algebras in \cite{G2, F4, D4}.

The area where usage of gradings has led to the most fruitful
results is construction of contractions of Lie algebras
\cite{normalizer_sl(3)_Cartan, HrNo, normalizer_sl(3)_Pauli}.
Apart from the classical (i.e.\ continuous) contractions, the
gradings of Lie algebras enabled to construct new contractions of
a dif\/ferent type. These contractions are called discrete, since
they cannot be obtained by means of continuous process.
Interesting results are also obtained when using the f\/ine gradings
for construction of deformations of the algebras.

The main results of that fundamental article by J.~Patera and
H.~Zassenhaus comprise a~statement about equivalence of the terms
grading and group grading, and a statement about one-to-one
correspondence between f\/ine gradings of f\/inite-dimensional simple
complex Lie algebras and MAD-groups. Therefrom, the crucial
consequence was derived that for description of all f\/ine gradings
of a f\/inite-dimensional simple complex Lie algebra it is
suf\/f\/icient to classify all the MAD-groups, which was done in the
article \cite{OLG2} in 1998. On the basis of this classif\/ication,
f\/ine gradings for several low-rank Lie algebras were found
\cite{complex_sp(4), complex_sl(4), complex_o(4)}.

A break-through moment has come when A.~Elduque revealed that the
proof in \cite{OLG1} of the one-to-one correspondence between f\/ine
gradings and MAD-groups was not correct. He gave an example in
\cite{Elduque} of a 16-dimensional complex non-simple Lie algebra
whose grading subspaces cannot be indexed by group neither
semigroup elements. It was just the coincidence of the terms
grading and group grading, on which the proof of the one-to-one
correspondence between the f\/ine gradings and the MAD-groups on
f\/inite-dimensional simple complex Lie algebras was based.
Therefore, the relation between the MAD-groups and the f\/ine
gradings remains an open problem. Nevertheless, the ef\/forts to
f\/ind a counterexample on a f\/inite-dimensional simple complex Lie
algebra that would contradict that statement (of one-to-one
correspondence) were unsuccessful so far.

In reaction to this revelation, the results in our articles
\cite{complex_sp(4), complex_sl(4), complex_o(4)} need to be
revised. We have proved a statement about one-to-one
correspondence between the MAD-groups and the f\/ine group gradings.
This statement holds for all f\/inite-dimensional (not only simple)
complex Lie algebras. The results in \cite{complex_sp(4),
complex_sl(4), complex_o(4)} thus remain valid when we replace the
term f\/ine grading by f\/ine group grading.

In parallel to complex Lie algebras, we have studied also f\/ine
gradings of real Lie algebras. For them, no statement about
one-to-one correspondence between the f\/ine gradings and the
MAD-groups has ever been asserted, and therefore the mistake
explained above has not af\/fected the results. The article
\cite{real_forms} describes some f\/ine group gradings of real forms
of the algebras $sl(4,\mathbb C)$, $o(4,\mathbb C)$, and
$sp(4,\mathbb C)$. We have developed several methods for
constructing these gradings, and we managed to show that there
exist f\/ine group gradings of real forms which are not generated by
any MAD-group of the respective real form.

In this whole work, we only devote our attention to f\/ine gradings,
but we do not investigate their coarsenings. In this context, we
appreciate the result of \cite{G2}, proving a theorem by use of
which all coarsenings of group gradings of f\/inite-dimensional
simple complex Lie algebras can be obtained.

\section{Gradings of Lie algebras}\label{gradings}

\subsection[Definition of a grading]{Def\/inition of a grading}\label{gradings-definition}

A {\em grading} of a Lie algebra $L$ is a decomposition $\Gamma$
of the vector space $L$ into vector subspaces $L_j \neq \{0\}$, $j
\in \mathcal J$ such that $L$ is a direct sum of these subspaces
$L_j$, and, for any pair of indices $j, k \in \mathcal J$, there
exists $l \in \mathcal J$ such that $\left[L_j, L_k \right]
\subseteq L_l$. We denote the grading by \[ \Gamma : L
=\bigoplus_{j \in \mathcal J} L_j. \] Clearly, for any Lie algebra
$L \neq \{0\}$, there exists the trivial grading $\Gamma : L = L$,
i.e. the Lie algebra is not split up at all. The opposite extreme
of splitting the Lie algebra into as many subspaces $L_j$ as
possible is more interesting and useful, though.

We call a grading $\widetilde{\Gamma} : L = \oplus_{j \in \mathcal
J} \oplus_{i \in \mathcal J_j} L_{ji}$ a {\em refinement} of the
grading $\Gamma : L = \oplus_{j \in \mathcal J} L_j$, when
$\oplus_{i \in \mathcal J_j} L_{ji} = L_j$ for each $j \in
\mathcal J$. A grading $\Gamma$ is called {\em fine} if each
ref\/inement $\widetilde{\Gamma}$ of $\Gamma$ is equal to $\Gamma$
itself (i.e.~$\Gamma$ does not have any proper ref\/inement
$\widetilde{\Gamma} \neq \Gamma$). A grading is necessarily f\/ine
when all its grading subspaces are one-dimensional. Such gradings
are called {\em finest}, and they have the practical ef\/fect of
def\/ining immediately a basis of the Lie algebra.

In the opposite direction, we can obtain any grading of $L$ from
some f\/ine grading of $L$, by merging some grading subspaces
together. Such process is called a {\em coarsening} of the
grading. The relations of ref\/inement and coarsening def\/ine an
ordering on the set of all gradings of a~given Lie algebra $L$.
Practically, we can illustrate the ordering in a hierarchy, whose
bottom nodes represent the f\/ine gradings of $L$ and whose top node
is the trivial grading (the whole Lie algebra~$L$ itself). If a
grading $\Gamma_{r}$ is connected in the hierarchy by an edge with
a grading $\Gamma_{s}$ on a lower level, it means that
$\Gamma_{s}$ is a ref\/inement of $\Gamma_{r}$, and that $\Gamma_r$
is a coarsening of $\Gamma_s$.

There are three special grading types (Cartan, Pauli, and
orthogonal), which are known on the inf\/inite series of classical
complex Lie algebras; Cartan gradings on all of the series $A_n$,
$B_n$, $C_n$, $D_n$, Pauli gradings and orthogonal gradings on
$A_n$ only:
\begin{itemize}\itemsep=0pt
    \item Cartan grading is the most notorious example. It is
    derived from the theory of root decomposition, and thus it is
    often referred to as the root grading.
    \item Pauli grading is a decomposition into powers of
    generalized matrices $P_m, Q_m \in \mathbb C^{m \times
    m}$, as introduced in \cite{PZpauli}:
    \[Q_m = \left(
        \begin{smallmatrix}
            0 & 1 & 0 & \ldots & 0\\
            0 & 0 & 1 & \ldots & 0\\
            \vdots & \vdots & \vdots & \ddots & \vdots\\
            0 & 0 & 0 & \ldots & 1\\
            1 & 0 & 0 & \ldots & 0
        \end{smallmatrix} \right), \quad P_m = \diag \left(1, \omega, \omega^2, \ldots,
    \omega^{m-1}\right), \quad \mathrm{where\ \ } \omega = \exp\left(\frac{2 \pi i}{m}\right). \]
    \item The orthogonal grading, introduced in \cite{kostrikin}, is
    referred to as orthogonal, since any pair of the grading subspaces
    is mutually orthogonal with respect to the scalar product def\/ined
    by $(A,B)=\tr(AB^+)$.
\end{itemize}
All gradings of these special types are f\/ine, with the only
exception of the Cartan grading on the non-simple algebra $D_2 =
o(4,\mathbb C)$.

Let us demonstrate the basic grading terminology on the well
explored algebra $sl(2,\mathbb C)$. This algebra has four
gradings; two of them are f\/ine, and even f\/inest ($\Upsilon_1$~--
the Cartan grading, $\Upsilon_2$~-- the Pauli grading coinciding
with the orthogonal grading), one is the trivial grading
$sl(2,\mathbb C)$, and the last one ($\Upsilon_0$) is neither
trivial, nor f\/ine:
\begin{gather*}
\Upsilon_0: sl(2,\mathbb C)=
    \Span^{\mathbb C} \left\{\left(\begin{smallmatrix}1&0\\0&-1\end{smallmatrix}\right)\right\} \oplus
    \Span^{\mathbb C} \left\{\left(\begin{smallmatrix}0&1\\0&0\end{smallmatrix}\right),
    \left(\begin{smallmatrix}0&0\\1&0\end{smallmatrix}\right)\right\},\\
\Upsilon_1: sl(2,\mathbb C)=
    \Span^{\mathbb C} \left\{\left(\begin{smallmatrix}1&0\\0&-1\end{smallmatrix}\right)\right\} \oplus
    \Span^{\mathbb C} \left\{\left(\begin{smallmatrix}0&1\\0&0\end{smallmatrix}\right)\right\} \oplus
    \Span^{\mathbb C}
    \left\{\left(\begin{smallmatrix}0&0\\1&0\end{smallmatrix}\right)\right\},\\
\Upsilon_2: sl(2,\mathbb C)=
    \Span^{\mathbb C} \left\{\left(\begin{smallmatrix}1&0\\0&-1\end{smallmatrix}\right)\right\} \oplus
    \Span^{\mathbb C} \left\{\left(\begin{smallmatrix}0&1\\1&0\end{smallmatrix}\right)\right\} \oplus
    \Span^{\mathbb C}
    \left\{\left(\begin{smallmatrix}0&1\\-1&0\end{smallmatrix}\right)\right\}.
\end{gather*}
The notion $\Span^{\mathbb{F}}\{M\}$ stands for the linear hull of
the set $M$ over the f\/ield $\mathbb{F}$.

\subsection{Equivalence of gradings}\label{gradings-equivalence}

The basic property of each automorphism $h \in {\rm Aut} \, L$
is that it preserves the commutation relations between the grading
subspaces $L_j$ of any grading $\Gamma$ of $L$: Clearly, for
$\left[L_j, L_k\right] \subseteq L_l$, we have $\left[h(L_j),
h(L_k)\right] = h(\left[L_j, L_k \right]) \subseteq h(L_l)$, and
thus the grading $\Gamma : L = \oplus_{j \in \mathcal J} L_j$
transformed by the automorphism $h \in {\rm Aut} \, L$ gives
rise to a new grading $\widetilde{\Gamma} : L = \oplus_{j \in
\mathcal J} h(L_j)$. This grading~$\widetilde{\Gamma}$ is
generally dif\/ferent from~$\Gamma$, however, its structure (meaning
the commutation relations between the grading subspaces,
dimensions of the grading subspaces, etc.) is the same as in~$\Gamma$. Therefore, from the structural perspective, the two
gradings $\widetilde{\Gamma}$ and $\Gamma$ equivalent. More
precisely, two gradings $\Gamma : L = \oplus_{j \in \mathcal J}
L_j$ and $\widetilde{\Gamma} : L = \oplus_{k \in \mathcal K} M_k$
are called {\em equivalent}, when there exist an automorphism $h
\in {\rm Aut} \, L$ and a bijection $\pi : \mathcal J \mapsto
\mathcal K$ such that $h(L_j) = M_{\pi (j)}$ for each $j \in
\mathcal J$. The equivalence is denoted by $\Gamma \cong
\widetilde{\Gamma}$.

\subsection{Group gradings}\label{group_gradings-definition}

Now we describe a specif\/ic type of a grading, namely a so-called
group grading. The most notorious case of a group grading is the
$\mathbb Z_2$-grading introduced by E.~In\"{o}n\"{u} and E.~Wigner
when decomposing a Lie algebra $L$ into two non-zero grading
subspaces $L_0$ and $L_1$, where \[[L_0, L_0]\subseteq L_0, \qquad
[L_0, L_1]\subseteq L_1, \qquad [L_1, L_1]\subseteq L_0.\]

A grading $\Gamma : L= \oplus_{j\in\mathcal J} L_j$ is called a
{\em group grading} if the index set $\mathcal J$ is a subset of
an Abelian group $G$ (whose binary operation is denoted by  $*$),
and, for any pair of indices $j,k\in\mathcal J$, it holds that
\begin{equation}\label{group-grading-commutation}
[L_j, L_k] \neq \{0\} \quad \Rightarrow \quad [L_j, L_k] \subseteq
L_{j*k}.
\end{equation}
Such group grading is often called a {\em $G$-grading} of the Lie
algebra $L$. If a ref\/inement of a group grading is again a group
grading, then we call it a {\em group refinement}.

For a $G$-grading of a f\/inite-dimensional Lie algebra $L$, we can
assume, without loss of generality, that the group $G$ is f\/initely
generated. If not, then $G$ can be replaced by its subgroup
generated by the elements of the f\/inite index set $\mathcal J$.

Two basic questions arise with respect to the group gradings:
\begin{itemize}\itemsep=0pt
    \item Does the group $G$ exist for each grading? In other
    words, is each grading also a group grading?
    \item In case the group $G$ exists, is it determined
uniquely for the respective group grading?
\end{itemize}
The latter question has an easy answer: the choice of the group
$G$ is not unique. However, there is one signif\/icant case among
all the possible index groups generated by $\mathcal J$, called
the {\em universal} (grading) {\em group}. It has the interesting
property (shown in \cite{G2}) that the index set of any coarsening
of the original grading is embedable into an image of the
universal group by some group epimorphism.

The question whether each grading is also a group grading seemed
to be positively answered in \cite{OLG1}. However, A.~Elduque in
\cite{Elduque} gave an example of a grading (on a 16-dimensional
complex non-simple algebra), whose grading subspaces cannot be
indexed by elements of any Abelian group neither semigroup while
satisfying the commutation relations
(\ref{group-grading-commutation}).

The distinction between group and semigroup is important in these
considerations. Another example of a grading whose grading
subspaces cannot be indexed by group elements is the Cartan-graded
$D_2 = o(4,\mathbb C)$, in which we further split the
two-dimensional Cartan subspace into two one-dimensional subspaces
while retaining the grading properties. This f\/ine grading cannot
be indexed by any group, but it can be indexed by a semigroup.
Analogous situation as in this example would repeat for any
semisimple non-simple complex Lie algebra, wherein we further
ref\/ine its non-f\/ine Cartan grading into a f\/ine grading, whose
grading subspaces can be indexed by a semigroup, but not by any
group.

It has been proved in \cite{G2} for gradings of simple complex Lie
algebras that if we embed the grading indices into a semigroup,
then they can be also embedded into a group.

\section{Automorphisms and group gradings} \label{automorphismsXgradings}

In this section we show how group gradings can be constructed by
means of automorphisms of the Lie algebra.

For a diagonalizable automorphism $g \in {\rm Aut}\,L$, the
eigensubspaces $L_j$ corresponding to dif\/ferent eigenvalues
$\lambda_j$ of the automorphism $g$ compose a group grading of the
algebra $L$. Indeed, for $X_j \in L_j$ and $X_k \in L_k$, we have
$g(\left[ X_j, X_k \right]) = \left[ g(X_j), g(X_k) \right] =
\left[ \lambda_j X_j, \lambda_k X_k \right] = \lambda_j \lambda_k
\left[ X_j, X_k  \right]$. Thus the commutator of $X_j$ and $X_k$
is either zero, or it lies in the eigensubspace of $g$
correspon\-ding to the eigenvalue $\lambda_j \lambda_k$. As $g$ is
diagonalizable, the sum of all its eigensubspaces makes up the
whole Lie algebra $L$, and therefore, the decomposition $\Gamma :
L = \oplus_j L_j$ into eigensubspaces~$L_j$ of $g$ is a grading of
$L$. We denote $\Gamma = \gr(g)$. Since each automorphism is a
non-singular mapping, the spectrum of~$g$ does not contain $0$.
Let us def\/ine $G$ as the smallest subgroup of the multiplicative
group $\mathbb C\setminus\{0\}$ such that $G$ contains the whole
spectrum $\sigma(g)$ of the automorphism~$g$. Then we can take the
spectrum of $g$ for the index set $\mathcal J\subseteq G$, thus
obtaining the group grading $\Gamma : L = \oplus_{j\in\mathcal J}
L_j$.

Now let us consider a set $\mathcal G = (g_p)_{p \in \mathcal P}$
of diagonalizable mutually commuting automorphisms $g_p \in
{\rm  Aut}\,L$; the set $\mathcal G$ can be f\/inite or inf\/inite.
We can f\/ind a basis of $L$ consisting of simultaneous eigenvectors
of all the automorphisms $g_p \in \mathcal G$. Taking a direct sum
of the simultaneous eigensubspaces of all the automorphisms $g_p
\in \mathcal G$, we obtain a group grading of $L$, which we denote
by $\gr(\mathcal G)$. The grading subspaces are indexed by
eigenvalues corresponding to the automorphisms $g_p$. Even if the
set $\mathcal G$ is inf\/inite, we can restrict ourselves to a
f\/inite number of elements $g_p \in \mathcal G$, $p = 1, \ldots,
\ell$, which is suf\/f\/icient for splitting the vector space into the
simultaneous eigensubspaces, as well as for indexing these
(grading) subspaces. (That is thanks to the f\/inite dimension of
$L$.) Again, the index set is a subset of the smallest subgroup
$G$ of the group $\left(\mathbb C\setminus\{0\}\right)^{\ell}$,
such that $G$ contains $\sigma(g_1) \times \cdots \times
\sigma(g_{\ell})$. In other words, $\gr (\mathcal G)$ has the form
$\gr (\mathcal G) = \Gamma : L = \oplus_{j^1 \in \mathcal J^1}
\cdots \oplus_{j^{\ell} \in \mathcal J^{\ell}} (L_{j^1}^1 \cap
\cdots \cap L_{j^{\ell}}^{\ell})$, where $\gr (g_p) = \Gamma^p : L
= \oplus_{j^p \in \mathcal J^p} L_{j^p}^p$ are the group gradings
of $L$ generated by the individual automorphisms $g_p$. For any
subset $\tilde{\mathcal G} \subset \mathcal G$, the group grading
$\gr (\mathcal G)$ is a group ref\/inement of the group grading $\gr
(\tilde{\mathcal G})$.

Having described how a group grading of a Lie algebra $L$ can be
obtained by means of a given set of automorphisms, let us now
approach the problem from the opposite direction, namely
investigate the automorphisms related to a given grading. Let
$\Gamma : L = \oplus_{j \in \mathcal J} L_j$ be a~grading of $L$
(not necessarily a group grading). The following notion will play
an important role in our considerations: \[
\Diag(\Gamma) = \{ g \in  {\rm Aut}\,L \, \vert\,
g/_{L_j} = \alpha_j {\rm Id} {\rm\ for \ any} \ j\in \mathcal J\}. \]

Directly from this def\/inition we can derive the following:
\begin{itemize}\itemsep=0pt
    \item An  automorphism  $g$  belongs to $ \Diag (\Gamma)$ if and
    only if $g(X) = \lambda_j X$ for all $X \in L_j$, $j \in \mathcal J$,
    where $\lambda_j \neq 0$ depends only on $g \in {\rm Aut}\,L$ and on
    $j \in \mathcal J$.
    \item $\Diag (\Gamma)$ is a subgroup  of $ {\rm Aut}\,L$, all automorphisms in
    $\Diag(\Gamma)$ are diagonalizable and mutually commute.
    \item For a group grading  $\Gamma = \gr(\mathcal G)$ generated by an arbitrary set
    $\mathcal G$ of mutually commuting diagonalizable automorphisms in ${\rm Aut}\, L$, it holds that $\Diag(\Gamma)\supseteq \mathcal G.$
    \item Let $\Gamma$ be a grading of $L$ (not necessarily a group
    grading). Then either $\gr(\Diag(\Gamma))=\Gamma$, or
    $\Gamma$ is a proper ref\/inement of $ \gr(\Diag(\Gamma))$.
\end{itemize}

There indeed do exist cases of such grading $\Gamma$ which is a
proper ref\/inement of $\gr(\Diag(\Gamma))$. The following theorem
proves, however, that this case cannot occur for group gradings of
f\/inite-dimensional complex Lie algebras, and, moreover, that all
group gradings of these algebras can be obtained by means of
automorphisms.

\begin{theorem}\label{group_gradings-automorphisms}
Let $\Gamma : L = \oplus_{j\in\mathcal J} L_j$ be a group grading
of a finite-dimensional complex Lie algebra. Then
\begin{itemize}\itemsep=0pt
    \item there exists a set of automorphisms $\mathcal G \subseteq {\rm Aut}\,
    L$ such that $\gr(\mathcal G) = \Gamma$, and
    \item $\gr(\Diag(\Gamma))=\Gamma$.
\end{itemize}
\end{theorem}

\begin{remark}
The fact that during the course of the proof of the Theorem
\ref{group_gradings-automorphisms} in \cite{PhD-thesis} we def\/ine
the set $\mathcal G$ in such a way that it is a group of
automorphisms does not mean that we necessarily need whole groups
of automorphisms for splitting the algebra into f\/ine group
gradings. On the contrary, in practice the set $\mathcal G$
fulf\/illing the role as stated in the Theorem
\ref{group_gradings-automorphisms} is a f\/inite set of
automorphisms containing just a very few elements.
\end{remark}

Let us now direct our attention to f\/ine group gradings. Remember
that we may ref\/ine the grading by enlarging the set of mutually
commuting diagonalizable automorphisms applied on the algebra.
That is why we make use of the term MAD-group introduced by
J.~Patera and H.~Zassenhaus: Let $\mathcal G$ be a subset of
${\rm Aut}\, L$ fulf\/illing the following properties:
\begin{itemize}\itemsep=0pt
    \item any $g\in \mathcal G$ is diagonalizable,
    \item $fg=gf$ for any $f,g \in \mathcal G$,
    \item if $h \in  {\rm Aut}\, L$ is diagonalizable and $hg=gh$ for any
$g\in \mathcal G$, then $h\in \mathcal G$.
\end{itemize}
Such $\mathcal G$ is called a {\em MAD-group} in ${\rm Aut}\,L$, which is an abbreviation  for maximal Abelian group of
diagonalizable automorphisms. (It follows obviously from the
def\/ining conditions imposed on the elements of $\mathcal G$ that
the word `group' in the notion is justif\/ied.)

\begin{theorem}\label{group_gradings-1-1}
Let $\Gamma : L = \oplus_{j\in\mathcal J} L_j$ be a group grading
of a finite-dimensional complex Lie algebra~$L$. Then $\Gamma$ is
a fine group grading if and only if the set $\Diag(\Gamma)$ is a
MAD-group in ${\rm Aut}\, L$.
\end{theorem}

\noindent It is this Theorem \ref{group_gradings-1-1} to which we
refer further in the text when talking about one-to-one
correspondence between f\/ine group gradings and MAD-groups of
f\/inite-dimensional complex Lie algebras. Both the Theorems
\ref{group_gradings-automorphisms} and \ref{group_gradings-1-1}
are proved in \cite{PhD-thesis}, or previously also in \cite{OV}.

\section[Methods for finding fine group gradings]{Methods for f\/inding f\/ine group gradings} \label{gradings-methods}

Our aim is to f\/ind f\/ine group gradings of the chosen Lie algebras,
which are either complex or real (real forms of complex Lie
algebras). In this process we use several methods, each of them
applicable for the various algebras in a dif\/ferent way, and also
with a dif\/ference in the strength of the result. In the sequel, we
list all the various methods used and specify their applicability
for the various Lie algebras in question.

\subsection{MAD-group method}\label{MAD-group_method}

As stated in Theorem \ref{group_gradings-1-1}, there is a
one-to-one correspondence between f\/ine group gradings of a~f\/inite-dimensional complex Lie algebra $L$ and MAD-groups of
automorphisms in ${\rm Aut}\, L$.

One can expect that some subgroups of ${\rm Aut}\,L$ would
generate equivalent gradings. That would of course be of no
interest to us, since we are looking for gradings with dif\/ferent
structural properties, i.e.\ non-equivalent. Luckily, there is an
easy key between equivalent f\/ine group gradings of $L$ and the
MAD-groups of automorphisms in ${\rm Aut}\, L$ that generate
them; and it uses the term of conjugate sets: Let $\mathcal G_1$
and $\mathcal G_2$ be subsets of ${\rm Aut}\,L$. We call these
subsets {\em conjugate} when there exists an automorphism $h \in
{\rm Aut}\,L$ such that $h \mathcal G_1 h^{-1} = \mathcal
G_2$, and we denote conjugate subsets by $\mathcal G_1 \cong
\mathcal G_2$.

\begin{theorem}\label{conj-equiv}
Let $\mathcal G_1, \mathcal G_2 \in {\rm Aut}\,L$ be
MAD-groups on a finite-dimensional complex Lie algebra~$L$; let
$\Gamma_1 = \gr(\mathcal G_1)$ and $\Gamma_2 = \gr(\mathcal G_2)$
be the fine group gradings of $L$ generated by $\mathcal G_1$ and
$\mathcal G_2$ respectively. The gradings $\Gamma_1$ and
$\Gamma_2$ are equivalent if and only if the MAD-groups $\mathcal
G_1$ and $\mathcal G_2$ are conjugate.
\end{theorem}

\begin{remark}
The equivalence in Theorem \ref{conj-equiv} is valid for f\/ine
group gradings and MAD-groups, but not for group gradings
generated by any sets of automorphisms in general. There we would
only have one implication in place, namely that conjugate subsets
$\mathcal G_1, \mathcal G_2 \subseteq {\rm Aut}\,L$ generate
equivalent group gradings $\gr(\mathcal G_1)\cong\gr(\mathcal
G_2)$. However, if we consider the sets $\mathcal G_i$ not as just
general subsets of ${\rm Aut}\,L$, but as the maximal sets of
automorphisms that leave the gradings $\gr(\mathcal G_i)$
invariant~-- i.e.~$\mathcal G_i = \Diag (\gr (\mathcal G_i))$, then
we end up with equivalence again; namely that the (f\/ine or
non-f\/ine) group gradings $\Gamma_i = \gr(\mathcal G_i)$ are
equivalent if and only if the sets $\mathcal G_i$ are conjugate.
\end{remark}

\noindent Now we come to the applicability of this `MAD-group'
method for the various Lie algebras in question (remember that we
keep on limiting ourselves only to the f\/inite-dimensional
classical complex Lie algebras and their real forms).

\subsubsection*{MAD-groups and f\/ine group gradings of the algebras
$\boldsymbol{A_n}$, $\boldsymbol{B_n}$, $\boldsymbol{C_n}$, $\boldsymbol{D_n}$} 

This is the most comfortable case, because, with the exception of
the algebra $D_4$, the MAD-groups were fully classif\/ied for all
the classical complex Lie algebras~\cite{OLG2}. Therefore, we
are able to f\/ind all the f\/ine group gradings, as explained in
Theorem~\ref{group_gradings-1-1}. Nevertheless, the main question,
which still remains unsolved, is whether this way leads to all the
possible f\/ine gradings.

The algebra $D_2$ is exceptional among the classical complex Lie
algebras, because it is not simple, but only semisimple. It has a
f\/ine group grading which is not f\/ine grading.

On simple complex Lie algebras, there has so far not been found
any f\/ine group grading, which would not be f\/ine grading at the
same time. That leaves open the hypothesis that for
f\/inite-dimensional simple complex Lie algebras the terms `grading'
and `group grading' could coincide.

\subsubsection*{MAD-groups and f\/ine group gradings of real forms
of the complex Lie algebras} 

For real algebras, we unfortunately do not possess any analogue to
the Theorems \ref{group_gradings-automorphisms} and
\ref{group_gradings-1-1}. Nevertheless, as well as for the
classical complex Lie algebras $A_n$, $B_n$, $C_n$, $D_n$, also for the
real forms of these complex algebras all the MAD-groups were
classif\/ied~\cite{OLG3}.

MAD-groups of the real forms can be obtained from MAD-groups of
the complex algebras as follows: Let $\mathcal G$ be a MAD-group
on the complex algebra $L$, and let $\mathcal G^{\mathbb R} =
\left\{ g \in \mathcal G \,\vert\, \sigma (g) \subset \mathbb R
\right\}$ be its so-called {\em real part}, namely the subgroup of
$\mathcal G$ containing all automorphisms with real spectrum. This
set $\mathcal G^{\mathbb R}$ is said to be maximal if there exists
no such MAD-group $\widetilde{\mathcal G}$ on $L$ (non-conjugate
to $\mathcal G$) that $\mathcal G^{\mathbb R}$ is conjugate to
some proper subgroup of $\widetilde{\mathcal G}^{\mathbb R}$. For
all the classical complex Lie algebras $L$, except for $D_4$, it
was proved in \cite{OLG3} that each MAD-group $\mathcal F$ on a
real form $L_{\mathbf J}$ of~$L$ is equal to the maximal real part
$\mathcal G^{\mathbb R}$ of some MAD-group $\mathcal G$ on $L$
restricted onto the real form~$L_{\mathbf J}$.

For general Lie algebras over $\mathbb R$ it has not been proved
that MAD-groups generate f\/ine group gradings. However, for real
forms of the classical complex Lie algebras, it follows from the
concrete construction of the MAD-groups that the gradings
generated by these MAD-groups are already f\/ine group gradings. The
opposite direction is not true, though; i.e.\ not all of the f\/ine
group gradings of the real forms are generated by MAD-groups of
these real forms, and not even on real forms of the classical
simple complex Lie algebras. In \cite{real_forms} we have found
several counterexamples, using some of the methods described
below. All of them correspond to the cases where the universal
group of the corresponding complex grading contains factors other
than $\mathbb Z$ or $\mathbb Z_2$ (i.e.\ it contains e.g.\ $\mathbb
Z_3$ or $\mathbb Z_4$ etc.); it is because the eigenvalues of
orders higher than 2 are not elements of $\mathbb R$.

\subsection{Displayed method}\label{displayed_method}

Searching for f\/ine group gradings by means of MAD-groups is a
laborious process. For the algebras $sl(m,\mathbb C)$ we do not
have any easier way, but for the other classical complex Lie
algebras $o(m,\mathbb C)$ and $sp(m,\mathbb C)$, which are
subalgebras of $sl(m,\mathbb C)$, we can use the so-called
`displayed' method. This method is applicable not only for the
complex Lie algebras $o(m, \mathbb C)$ and $sp(m, \mathbb C)$, but
also for their real forms (which are subalgebras of the real forms
of $sl(m,\mathbb C)$).

The method consists in the following principle: A {\em subalgebra}
$o_K(m, \mathbb C)$ or $sp_K(m, \mathbb C)$ of the Lie algebra
$sl(m, \mathbb C)$ is said to be {\em displayed by a grading}
$\Gamma$ of $sl(m, \mathbb C)$, when $o_K(m, \mathbb C)$ or
$sp_K(m, \mathbb C)$ respectively is equal to a direct sum of
selected grading subspaces from $\Gamma$. Then, it necessarily
holds that such a direct sum is also a grading of $o_K(m, \mathbb
C)$ or $sp_K(m,\mathbb C)$ respectively.

It follows from the classif\/ication of MAD-groups on the classical
complex Lie algebras that a f\/ine grading $\Gamma$ of $sl(m,\mathbb
C)$ displays $o_K(m,\mathbb C)$ or $sp_K(m,\mathbb C)$ if and only
if the MAD-group $\mathcal G = \Diag(\Gamma)$ contains an outer
automorphism ${\rm Out}_K$ with $K=K^T$ or $K=-K^T$ respectively. Let us
explain this statement on the case of $o(m,\mathbb C)$; the case
of $sp(m,\mathbb C)$ would be just analogous.

\begin{itemize}\itemsep=0pt
    \item Having the outer automorphism ${\rm Out}_K \in \mathcal G$
    with $K = K^T$, the eigensubspace of ${\rm Out}_K$ corresponding to
    the eigenvalue $+1$ is the subalgebra $o_K(m,\mathbb C)$.
    \item Any other element $g \in \mathcal G$ can then be
    restricted to $o_K(m,\mathbb C)$ while preserving the $\mathbb
    Z_2$-grading generated by ${\rm Out}_K$, because $g$ and ${\rm Out}_K$
    commute.
    \item The set $\{g \vert _{o_K(m,\mathbb C)} \, \vert \, g \in
    \mathcal G \}$ is a MAD-group on $o_K(m,\mathbb C)$, since
    every automorphism on $o_K(m, \mathbb C)$ can be extended to
    an automorphism of $sl(m,\mathbb C)$ commuting with ${\rm Out}_K$.
    \item We can express the MAD-group $\mathcal G$ on
    $sl(m,\mathbb C)$ in the form $\mathcal G = \mathcal H \cup
    {\rm Out}_K \mathcal H$, where $\mathcal H$ is the set of all inner
    automorphisms in $\mathcal G$. The corresponding MAD-group on
    $o_K(m, \mathbb C)$ then has the form $\{ g \vert
    _{o_K(m,\mathbb C)} \, \vert \, g \in \mathcal G \} = \{ g
    \vert _{o_K(m,\mathbb C)} \, \vert \, g \in \mathcal H \} $.
    Thus, the subgroup $\mathcal H$, upon restriction onto $o_K(m,\mathbb
    C)$, has the same splitting ef\/fect on the subalgebra $o_K(m,\mathbb
    C)$ as the original group $\mathcal G$.
\end{itemize}

The `displayed' method gives the same result in producing the f\/ine
group gradings as the `MAD-group' method in cases of the complex
subalgebras of $sl(m,\mathbb C)$; in other words, we obtain all
the f\/ine group gradings for these complex algebras.

For real forms of $o(m,\mathbb C)$ and $sp(m,\mathbb C)$, the
`displayed' method in fact turns into making an intersection of
the complex subalgebra $o(m,\mathbb C)$ and $sp(m,\mathbb C)$
respectively with a f\/ine group grading of the relevant real form
of $sl(m,\mathbb C)$. This is a consequence of the fact that real
forms of the subalgebras $o(m,\mathbb C)$ and $sp(m,\mathbb C)$ of
the complex algebra $sl(m,\mathbb C)$ are subalgebras of the real
forms of $sl(m,\mathbb C)$.

The gradings of the real forms of the subalgebras obtained by this
`displayed' method are f\/ine group gradings, however, we have no
certainty of getting all the f\/ine group gradings of the respective
real forms. Nevertheless, this method is still quite powerful and
provides a high number of f\/ine group gradings of the real forms.

\subsection{Fundamental method}\label{fundamental_method}

As can be seen from above, the `MAD-group' method and the
`displayed' method are excellent when searching for f\/ine group
gradings of the complex Lie algebras, but not so strong in case of
the real forms. Thus, we still need to broaden our scope by
another method, which is specialized on (and applicable only for)
the real forms. We call this method `fundamental', because it
derives directly from the def\/inition of the real form. It assumes
that we already dispose of the f\/ine group gradings of the
respective complex algebra.

\begin{theorem}\label{fundamental}
Let $\Gamma:L=\oplus_{j\in\mathcal J}L_j$ be a fine group grading
of a classical complex Lie algebra $L$. Let $\mathbf J$ be an
involutive antiautomorphism on $L$, let $L_{\mathbf J}$ be the
real form of $L$ defined by $\mathbf J$, and let $Z_{j,l}$ be
elements of $L_j$ fulfilling the following properties:
\begin{itemize}\itemsep=0pt
    \item $(Z_{j,1},\ldots,Z_{j,l_j})$ is a basis of the grading
    subspace $L_j$ for each $j\in\mathcal J$; i.e.
    $L_j = \mathrm{span}^{\mathbb C}(Z_{j,1},$ $\ldots,Z_{j,l_j})$; and
    \item $\mathbf J(Z_{j,l})=Z_{j,l}$; i.e.\ $Z_{j,l}\in L_{\mathbf J}$ for all
    $Z_{j,l}\in L_j$ for each $j\in\mathcal J$.
\end{itemize}
Then the decomposition $\Gamma^{\mathbf J}:L_{\mathbf
J}=\oplus_{j\in\mathcal J}L^{\mathbb R}_j$ into subspaces
$L^{\mathbb R}_j = \mathrm{span}^{\mathbb
R}(Z_{j,1},\ldots,Z_{j,l_j})$ is a fine group grading of the real
form $L_{\mathbf J}$.

 We then say that the fine group grading
$\Gamma$ of the complex Lie algebra $L$ {\em determines} the fine
group grading $\Gamma^{\mathbf J}$ of the real form $L_{\mathbf
J}$.
\end{theorem}

This method in practice turns into rather laborious calculations,
namely into looking for a~suitable antiautomorphism $\mathbf J$
and a suitable basis of the complex algebra $L$, such that the
basis vectors form not only the bases of the grading subspaces
$L_j$ of the complex grading $\Gamma$ of~$L$, but also lie in (and
thus form the basis of) the real form $L_{\mathbf J}$.

Not even this method ensures f\/inding all the f\/ine group gradings
of the real forms, however, we at least get as much as we can from
the complex f\/ine group gradings, and it is the strongest method we
dispose of for the real forms. This `fundamental' method as well
as the `real-basis' method, which follows below, were in detail
described and proved in~\cite{Ing-thesis}.

\subsection{Real basis method}\label{real_basis_method}

Finally, we describe a simplif\/ied version of the `fundamental'
method, called the `real basis' method. It applies only for the
real forms of the complex Lie algebras $sl(m,\mathbb C)$, it is
less powerful than the `fundamental' method, but, on the other
hand, much easier for practical application.

\begin{theorem}\label{real_basis}
Let $\mathcal G$ be a MAD-group on the complex Lie algebra
$sl(m,\mathbb C)$ and let $\Gamma : sl(m,\mathbb C) =
\oplus_{j\in\mathcal J}L_j$ be the fine group grading of
$sl(m,\mathbb C)$ generated by $\mathcal G$, such that all the
subspaces $L_j = \mathrm{span}^{\mathbb
C}(X_{j,1},\ldots,X_{j,l_j})$ have real basis vectors $X_{j,l} \in
sl(m,\mathbb R)$. Let $h$ be an automorphism in ${ \rm Aut}\,sl(m,\mathbb C)$, such that $\mathbf J=\mathbf J_0h$, where
$\mathbf J_0$ acts as complex conjugation on elements of
$sl(m,\mathbb C)$, is an involutive antiautomorphism on
$sl(m,\mathbb C)$. Then the fine group grading $\Gamma$ of
$sl(m,\mathbb C)$ determines a fine group grading of the real form
$L_{\mathbf J} = L_{\mathbf J_0 h}$ if and only if the
automorphism $h$ is an element of the MAD-group $\mathcal G$.
\end{theorem}

The simplicity of using this method follows from the fact that we
do not have to occupy ourselves with the elements of the Lie
algebra, but we only investigate the MAD-group $\mathcal G$, in
order to f\/ind out whether it contains a convenient automorphism
$h$. Obviously, the set of solutions provided by this method is
generally not complete.

 Throughout the task to f\/ind the f\/ine group gradings of
complex Lie algebras and of their real forms, we use the above
mentioned methods alternatively, depending on their applicability
and strength for the respective algebra. The concrete results are
summarized in Section~\ref{results}.

\section{Results}\label{results}

As announced earlier, our main aim was to f\/ind f\/ine group gradings
of certain classical complex Lie algebras and their real forms.
Naturally, one proceeds from those with the lowest ranks, since
low rank implies low dimension of the algebra, and thus (relative)
simplicity in the calculations. And last but not least argument is
the fact that the low-rank Lie algebras are most widely used in
the practice of physics.

On $A_1=sl(2,\mathbb C)$ the two f\/ine group gradings (Cartan and
Pauli) have been known for long time already. The four f\/ine group
gradings of the Lie algebra $A_2 = sl(3,\mathbb C)$ were f\/irstly
published in \cite{formysl3c}, together with the f\/ine group
gradings of its three real forms.

In our work from the area of f\/ine gradings, we have naturally
started with the lowest rank where the results were missing,
namely rank two and algebras $B_2$, $C_2$, and $D_2$. The two
algebras $B_2 = o(5, \mathbb C)$ and $C_2 = sp(4, \mathbb C)$ are
isomorphic, and thus their grading properties are the same. Then
we continue with the algebra $D_2 = o(4, \mathbb C)$, the only
non-simple classical complex Lie algebra. Lastly, we move to the
algebras of rank three, namely two isomorphic algebras $A_3 =
sl(4, \mathbb C)$ and $D_3 = o(6, \mathbb C)$.

We have tried to combine all the various methods described in
Section \ref{gradings-methods}, in order to f\/ind as many f\/ine
group gradings as possible. The isomorphisms $B_2 \cong C_2$ and
$A_3 \cong D_3$ are not the only auxiliary relations we dispose
of. Additionally, we were able to benef\/it from the fact that $C_2
= sp(4, \mathbb C)$ and $D_2 = o(4, \mathbb C)$ are subalgebras of
$A_3 = sl(4, \mathbb C)$. That allows us to use the `displayed'
method, which is especially ef\/fective in the case of real forms.

The results were published in a series of articles
\cite{complex_sp(4), complex_sl(4), complex_o(4), real_forms}. Let
us recall that, in all these works, the lists of f\/ine group
gradings of the complex Lie algebras are complete (according to
our Theorem \ref{group_gradings-1-1}), whereas for real forms,
where no such statement has been proved, we cannot claim our
results to be exhaustive solutions to the problem of f\/ine group
gradings.

\subsubsection*{$\boldsymbol{B_2=o(5,\mathbb C)}$, $\boldsymbol{C_2=sp(4,\mathbb C)}$}
\label{gradings-B2+C2}

These (mutually isomorphic) algebras are 10-dimensional. There are
three non-conjugate MAD-groups on the complex algebras, and thus
three non-equivalent f\/ine group gradings, which were found by the
`MAD-group' method in \cite{complex_sp(4)} and then conf\/irmed for
$C_2 = sp(4, \mathbb C)$ also by the `displayed' method in
\cite{complex_sl(4)} (where several representations with dif\/ferent
def\/ining matrices $K$ of $sp_K(4, \mathbb C)$ appear).

Through coincidence, the number of real forms of these algebras is
three, too. The detailed results on the real forms are in both the
articles \cite{complex_sp(4)} and \cite{real_forms}; found by
means of the `MAD-group' method in the former and by means of the
`displayed' method in the latter. Both of the methods by
def\/inition have to provide the same result, only in dif\/ferent
representations. It is only by mistake that the Cartan f\/ine group
grading of $usp(2,2$) is missing in~\cite{complex_sp(4)}~-- the
respective MAD-group was forgotten there; hence the result in
\cite{real_forms} is richer by this one f\/ine group grading.

\begin{table}[t]\centering
\caption{Fine group gradings of $o(5,\mathbb C) \cong sp(4,\mathbb
C)$ and of their real forms.}\label{gradings-B2+C2-table}
\vspace{1mm}

\begin{tabular}{|c|l|} \hline \hline
 \multicolumn{2}{|l|}{complex algebras $B_2 \cong C_2$}\\ \hline
  & $\Gamma_1$: $1 \times 2$-dim + $8 \times 1$-dim (Cartan) \\
 $o(5, \mathbb C) \cong sp(4, \mathbb C)$ & $\Gamma_2$: $10 \times 1$-dim \\
  & $\Gamma_3$: $10 \times 1$-dim \\ \hline \hline
 \multicolumn{2}{|l|}{real forms}\\ \hline
 \tsep{1ex}$so(5,0) \cong usp(4,0)$ & $\Gamma_3^{\mathbb R}$ \\
 $so(4,1) \cong usp(2,2)$ & $\Gamma_1^{\mathbb R}$, $\Gamma_2^{\mathbb R}$ , $\Gamma_3^{\mathbb R}$ \\
 $so(3,2) \cong sp(4, \mathbb R)$ & $\Gamma_1^{\mathbb R}$, $\Gamma_2^{\mathbb R}$ , $\Gamma_3^{\mathbb R}$ \\ \hline \hline
\end{tabular}
\end{table}

\subsubsection*{$\boldsymbol{D_2=o(4,\mathbb C)}$}\label{gradings-D2}

The six-dimensional algebra $D_2 = o(4, \mathbb C)$ is the only
non-simple case among the classical complex Lie algebras. It has
six non-conjugate MAD-groups, one of them (Cartan) generating a~f\/ine group grading, which is not f\/ine grading. It splits the
complex algebra into four one-dimensional grading subspaces and
one two-dimensional subspace (the Cartan subalgebra). On simple
classical complex Lie algebras of rank $r$, the $r$-dimensional
Cartan subspace cannot be decomposed any further while preserving
the grading properties within the Cartan grading. But the
non-simple algebra $D_2 = o(4,\mathbb C)$ is composed of two
instances of $sl(2,\mathbb C)$, and a direct sum of two instances
of the Cartan-graded $sl(2,\mathbb C)$ is a f\/ine grading of
$o(4,\mathbb C)$, which is a non-group ref\/inement of the group
grading generated by the Cartan MAD-group on $o(4,\mathbb C)$.

Note that also another two of the f\/ine group gradings of
$o(4,\mathbb C)$ are composed of two f\/ine group gradings of the
algebra $sl(2,\mathbb C)$, one is made up of two instances of the
Pauli-graded $sl(2,\mathbb C)$, and the other one consists of one
Cartan-graded $sl(2,\mathbb C)$ and one Pauli-graded $sl(2,\mathbb
C)$. The remaining three f\/ine group gradings of $o(4,\mathbb C)$
cannot be expressed in terms of f\/ine group gradings of
$sl(2,\mathbb C)$. The full list of f\/ine group gradings of the
complex algebra $D_2$ are in \cite{complex_o(4)}.

The algebra $D_2 = o(4,\mathbb C)$ has four real forms and we
derive their f\/ine group gradings by means of the `displayed'
method from the real forms of $sl(4,\mathbb C)$, this set of
solutions (provided in \cite{real_forms}) is not proved to be
exhaustive, though.

\begin{table}[t]\centering
\caption{Fine group gradings of $o(4,\mathbb C)$ and of its real
forms.} \label{gradings-D2-table}
\vspace{1mm}

\begin{tabular}{|c|l|} \hline \hline
 \multicolumn{2}{|l|}{complex algebra $D_2$}\\ \hline
  & $\Gamma_1$: $1 \times 2$-dim + $4 \times 1$-dim (Cartan $\times$ Cartan) \\
 $o(4, \mathbb C)$ & $\Gamma_2$: $6 \times 1$-dim (Cartan $\times$ Pauli) \\
  & $\Gamma_3$: $6 \times 1$-dim (Pauli $\times$ Pauli) \\
  & $\Gamma_4$: $6 \times 1$-dim \\
  & $\Gamma_5$: $6 \times 1$-dim \\
  & $\Gamma_6$: $6 \times 1$-dim \\ \hline \hline
 \multicolumn{2}{|l|}{real forms}\\ \hline
 \tsep{1ex}$so^*(4)$ & $\Gamma_2^{\mathbb R}$, $\Gamma_3^{\mathbb R}$, $\Gamma_6^{\mathbb R}$ \\
 $so(4,0)$ & $\Gamma_3^{\mathbb R}$, $\Gamma_4^{\mathbb R}$ \\
 $so(3,1)$ & $\Gamma_4^{\mathbb R}$, $\Gamma_5^{\mathbb R}$, $\Gamma_6^{\mathbb R}$ \\
 $so(2,2)$ & $\Gamma_1^{\mathbb R}$, $\Gamma_2^{\mathbb R}$, $\Gamma_3^{\mathbb R}$,
  $\Gamma_4^{\mathbb R}$, $\Gamma_5^{\mathbb R}$, $\Gamma_6^{\mathbb R}$ \\ \hline \hline
\end{tabular}
\end{table}

\begin{table}[t]\centering
\caption{Fine group gradings of $sl(4,\mathbb C) \cong o(6,\mathbb
C)$ and of their real forms.} \label{gradings-A3+D3-table}
\vspace{1mm}

\begin{tabular}{|c|l|} \hline \hline
 \multicolumn{2}{|l|}{complex algebras $A_3 \cong D_3$}\\ \hline
  & $\Gamma_1$: $1 \times 3$-dim + $12 \times 1$-dim (Cartan) \\
 $sl(4, \mathbb C) \cong o(6, \mathbb C)$ & $\Gamma_2$: $15 \times 1$-dim (Pauli) \\
  & $\Gamma_3$: $1 \times 3$-dim + $12 \times 1$-dim (orthogonal) \\
  & $\Gamma_4$: $1 \times 2$-dim + $13 \times 1$-dim \\
  & $\Gamma_5$: $1 \times 2$-dim + $13 \times 1$-dim \\
  & $\Gamma_6$: $15 \times 1$-dim \\
  & $\Gamma_7$: $15 \times 1$-dim \\
  & $\Gamma_8$: $1 \times 2$-dim + $13 \times 1$-dim \\ \hline \hline
 \multicolumn{2}{|l|}{real forms}\\ \hline
 \tsep{1ex}$sl(4, \mathbb R) \cong so(3,3)$ & $\Gamma_1^{\mathbb R}$, $\Gamma_3^{\mathbb R}$, $\Gamma_4^{\mathbb R}$,
  $\Gamma_5^{\mathbb R}$, $\Gamma_6^{\mathbb R}$, $\Gamma_7^{\mathbb R}$, $\Gamma_8^{\mathbb R}$ \\
 $su^*(4) \cong so(5,1)$ & $\Gamma_6^{\mathbb R}$, $\Gamma_7^{\mathbb R}$, $\Gamma_8^{\mathbb R}$ \\
 $su(4,0) \cong so(6,0)$ & $\Gamma_3^{\mathbb R}$, $\Gamma_7^{\mathbb R}$ \\
 $su(3,1) \cong so^*(6)$ & $\Gamma_2^{\mathbb R}$, $\Gamma_3^{\mathbb R}$, $\Gamma_4^{\mathbb R}$, $\Gamma_8^{\mathbb R}$\\
 $su(2,2) \cong so(4,2)$ & $\Gamma_2^{\mathbb R}$, $\Gamma_3^{\mathbb R}$, $\Gamma_4^{\mathbb R}$,
  $\Gamma_5^{\mathbb R}$, $\Gamma_6^{\mathbb R}$, $\Gamma_7^{\mathbb R}$, $\Gamma_8^{\mathbb R}$ \\ \hline \hline
\end{tabular}
\end{table}

\subsubsection*{$\boldsymbol{A_3=sl(4,\mathbb C)}$, $\boldsymbol{D_3=o(6,\mathbb C)}$}
\label{gradings-A3+D3}

The last algebra whose f\/ine gradings we have investigated is
already of rank three and of dimension f\/ifteen. We deal in fact
again with two algebras that are isomorphic, namely $A_3 = sl(4,
\mathbb C)$ and $D_3 = o(6, \mathbb C)$. The complex algebra has
eight non-conjugate MAD-groups, and thus eight non-equivalent f\/ine
group gradings~\cite{complex_sl(4)}.

The number of real forms of the algebra $sl(4,\mathbb C)$ is f\/ive,
and again, we also try to deliver as many f\/ine group gradings as
possible for them. We apply all the methods we dispose of as
explained in Section \ref{gradings-methods}, starting from the
easiest `MAD-group' method, continuing with the `real basis'
method, and, lastly, turning to the `fundamental' method. Each of
these methods brings in additional results (see~\cite{real_forms}), and those then enable, via the `displayed'
method, to obtain the richest possible results for the real forms
of the subalgebras $sp(4,\mathbb C)$ and $o(4,\mathbb C)$.

\section{Concluding remarks}\label{concl}

Let us conclude by the open problems and hypotheses which need to
be subject of further studies:

\begin{itemize}\itemsep=0pt
    \item The main question is the relationship between gradings
    and group gradings. Our hypo\-thesis is that for f\/inite-dimensional
    simple complex Lie algebras these two terms coincide.
    \item For the real forms of complex Lie algebras, it is to be
    clarif\/ied whether the `fundamental' method provides all the
    f\/ine group gradings. No counterexample has been found so far
    against this assumption.
\end{itemize}

\subsection*{Acknowledgements}

The author acknowledges f\/inancial support by N\v{C}LF (Nadace
\v{C}esk\'y liter\'arn\'{\i}
    fond) and  by the grants LC06002 and  MSM6840770039 of the
Ministry of Education, Youth, and Sports of the Czech Republic

\pdfbookmark[1]{References}{ref}

\LastPageEnding

\end{document}